\documentclass[reqno]{amsart}
\usepackage{amsfonts,amsmath,amssymb}

\newcounter{mnotecount}[section]

\newcommand{\rmnote}[1]{}

\numberwithin{equation}{section}

\newtheorem{thm}{Theorem}[section]

\newtheorem{cor}{Corollary}[section]

\begin{document}
\title[Einstein--scalar field Lichnerowicz equations]
{A variational analysis of Einstein--scalar field Lichnerowicz 
equations on compact Riemannian manifolds}
\author{Emmanuel Hebey}
\address{Emmanuel Hebey, Universit\'e de Cergy-Pontoise, 
D\'epartement de Math\'ematiques, Site de 
Saint-Martin, 2 avenue Adolphe Chauvin, 
95302 Cergy-Pontoise cedex, 
France}
\email{Emmanuel.Hebey@math.u-cergy.fr}
\author{Frank Pacard}
\address{Frank Pacard, Universit\'e Paris XII, D\'epartement de Math\'ematiques, 
61 avenue du G\'en\'eral de Gaulle, 94010 Cr\'eteil cedex, France}
\email{pacard@univ-paris12.fr}
\author{Daniel Pollack}
\address{Daniel Pollack, University of Washington, Department of Mathematics, Box 354350, Seattle, 
WA 98195--4350, USA}
\email{pollack@math.washington.edu}

\date{February 2, 2007}

\begin{abstract}
We establish new existence and non-existence results for positive solutions of the Einstein--scalar field
Lichnerowicz equation on compact manifolds.  This  equation arises from the Hamiltonian constraint 
equation for the Einstein--scalar field system in general relativity. Our analysis introduces variational 
techniques, in the form of the mountain pass lemma, to the analysis of the Hamiltonian constraint equation,
which has been previously studied by other methods.
\end{abstract}

\maketitle

\section{Introduction}\label{Intro}

One of the foundations in the mathematical analysis of the Einstein field equations of general relativity is the rigorous
formulation of the Cauchy problem. The basic local existence result of Choquet-Bruhat \cite{FB52}, and the important  
extension of this due to Choquet-Bruhat and Geroch \cite{CBG}, allows one to approach the study of 
globally hyperbolic spacetimes via the analysis of initial data sets.  The Gauss and Codazzi equations impose
constraints on the choices of initial data in general relativity, and these constraints are expressed by 
the Einstein constraint equations. This perspective, originally
studied in the context of vacuum spacetimes, has also been successfully employed in the study of many non-vacuum models
obtained by minimally coupling gravity to many of the classical matter and field sources, such as electromagnetism 
(via the Maxwell equations), Yang-Mills fields, fluids, and others  \cite{CBY, I95, IMaxP}.
One of the simplest non-vacuum systems is the Einstein-scalar field system which arises in coupling gravity to a scalar
field satisfying a linear or non-linear wave equation with respect to the Lorentz metric describing the
gravitational field.  The Einstein-scalar field system, when posed in this generality, includes as special cases the 
(massless or massive) Einstein-Klein-Gordon equations as well as the vacuum Einstein equations with a 
(positive or negative) cosmological constant.

\medskip

Einstein--scalar field theories have been the subject of interesting developments in recent years. Among these 
are the recent attempts to use such theories to explain the observed acceleration of the expansion of the universe
\cite{Ren1, Ren2, Ren3, Sa}. 
Using the conformal method, Choquet-Bruhat, Isenberg, and Pollack \cite{CBIP, CIP} reformulated the constraint 
equations for the Einstien--scalar field system  as a determined system of nonlinear partial differential equations. 
The equations are semi-decoupled in the constant mean curvature (CMC) setting. One of these equations, 
the conformally formulated momentum constraint, is a linear elliptic equation and its solvability is easy to address. 
The other one, the conformally formulated Hamiltonian constraint, is a nonlinear elliptic equation
(the Einstein--scalar field Lichnerowicz equation)
as in \eqref{EinLicEqt} below (see \cite{BI04} for a survey on the constraint equations, and in particular,
the conformal method). This nonlinear equation, which contains both a positive critical Sobolev nonlinearity 
and a negative power nonlinearity, turns out to be of great mathematical interest. 
In this paper we  provide a variational analysis of this equation under certain conditions on its coefficients.
The analysis of the Lichnerowicz equations which arise as the conformally formulated Hamiltonian constraint
equations in both vacuum and non-vacuum settings has, in the past, been conducted primarily by either the
method of sub and supersolutions (i.e. a barrier method) or by perturbation or fixed point methods.  This 
approach has been sufficient to allow for a complete understanding of solvability in, for example, the case 
of constant mean curvature vacuum initial data on compact manifolds \cite{I95}.  In \cite{CIP} this method was
applied to constant mean curvature initial data for the Einstein-scalar field system on compact manifolds. 
In a number of cases, the method of sub and supersolutions was shown to be sufficient to completely analyze the
solvability of the Einstein--scalar field Lichnerowicz equation. In other cases, the limitations of this 
method were exposed and only partial results were obtained.
We establish here two general theorems concerning non-existence and existence respectively, of positive 
solutions to the  Einstein--scalar field Lichnerowicz equation \eqref{EinLicEqt}.  These results are of interest due 
both to their application to questions of existence and non-existence of solutions of the Einstein--scalar field
constraint equations, as well as, more generally, the introduction of variational techniques to the analysis
of the constraint equations.
We expect that similar variational techniques will be of use in resolving other open questions concerning
initial data for the Cauchy problem in general relativity.

\medskip

In what follows we let $(M,g)$ be a smooth compact Riemannian manifold 
of dimension $n \ge 3$. We let also $H^1(M)$ be  the Sobolev space of functions in $L^2(M)$ with 
one derivative in $L^2(M)$. 
The $H^1$--norm on $H^1(M)$ is given by
$$\Vert u\Vert_{H^1} = \sqrt{\int_M\left(\vert\nabla u\vert^2+u^2\right)dv_g}
\hskip.1cm .$$
Let $2^\star = \frac{2n}{n-2}$, so that $2^\star$ is the critical Sobolev exponent 
for the embedding of $H^1$ into Lebesgue's spaces.  Let also $h$, $A$, and $B$ be smooth functions on $M$. We consider 
the following Einstein--scalar field Lichnerowicz type equations
\begin{equation}\label{EinLicEqt}
\Delta_gu + hu = Bu^{2^\star-1} + \frac{A}{u^{2^\star +1}}\hskip.1cm ,
\end{equation}
where $\Delta_g = -\hbox{div}_g\nabla$ is the Laplace-Beltrami operator, and $u > 0$. 
Unless otherwise stated, solutions are always required to be smooth and positive. 

\medskip

The relationship between the coefficients in \eqref{EinLicEqt} and initial data for the 
Einstein--scalar field system are as follows (see \cite{CIP} for more details).  We first note that the 
sign convention for the Laplace-Beltrami operator which we use here is  the opposite of the one used 
in \cite{CIP}.  The conformal initial data for the purely gravitational portion of the Einstein--scalar field
system consists of a background Riemannian metric $g$ (indicating a choice of conformal class for
the physical metric) together with a symmetric $(0,2)$-tensor 
$\sigma$ which is divergence-free and trace-free with respect to $g$ (so that $\sigma$ is
what is commonly referred to as a transverse-traceless, or TT-tensor) and a scalar function $\tau$ 
representing the mean curvature of the Cauchy surface $M$ in the  spacetime development
of the initial data set. The initial data for the scalar field consists of two functions, $\psi$ and  $\pi$ on $M$,
representing respectively the initial value for the scalar field and its normalized time derivative.  With respect to 
this set of conformal initial data, the constraint equations for the Einstein--scalar field system can be 
realized as a determined elliptic system whose unknowns consist of a positive scalar function $\phi$ 
and a vector field $W$ on $M$.  As previously remarked, in the CMC case (when $\tau$ is constant)
this system becomes semi-decoupled.  This means that the portion of it corresponding to the momentum
constraint equation is a linear, elliptic, vector equation for $W$ in which  the unknown $\phi$ does not
appear.  This equation has a unique solution when $(M, g)$ has no conformal Killing vector fields.  
The solution, $W$, of this ``conformally formulated momentum constraint equation" then appears in the
one of the coefficients of the ``conformally formulated Hamiltonian constraint equation" which is what we refer to 
as the Einstein--scalar field Lichnerowicz equation.  A positive solution
$\phi$ of the  Einstein--scalar field Lichnerowicz equation is then used with the vector field $W$ to
transform the ``conformal" initial data set $(g, \sigma, \tau, \psi, \pi)$ into a ``physical" initial data set 
satisfying the Einstein--scalar field constraint equations (see \cite{CIP}).  In terms of the conformal
initial data set and the vector field $W$ (satisfying the conformally formulated momentum constraint equation)
the coefficients of the Einstein--scalar field Lichnerowicz equation \eqref{EinLicEqt}  are
$$h=c_n\left(R(g)-|\nabla{\psi}|^2_{g}\right), \qquad
A=c_n\left(|\sigma + {\mathcal D} W|^2_{g}+\pi^2\right)$$
and
$$B=-c_n\left(\frac{n-1}{n} \tau^2 -4V(\psi) \right)$$
where  $c_n = \frac{n-2}{4(n-1)}$, $R(g)$ is the scalar curvature,
$\nabla$ is the covariant derivative for  $g$,
$V(\cdot)$ is the potential in the wave equation for the scalar field, 
and  the operator $\mathcal D$ is  the conformal Killing operator
relative to $g$,  defined by $({\mathcal D}W)_{ab} := \nabla_a W_b+ \nabla_b W_a -\frac{2}{n} g_{ab}\nabla_m W^m$.
The  kernel of $\mathcal D$  consists of the conformal Killing fields on $(M,g)$.  Note that relative to the 
notation of \cite{CIP}, we have $h={\mathcal R}_{g, \psi}$, $B=-{\mathcal B}_{\tau, \psi}$ 
and $A= {\mathcal A}_{g, W, \pi}$.

\medskip

We assume in what follows that $A \ge 0$ in $M$.  
This assumption implies no physical restrictions since
we always have that $A \ge 0$ in the original Einstein--scalar field theory.  One of the results 
of \cite{CIP} is the definition of a conformal invariant, the Yamabe--scalar field conformal invariant, whose
sign can be used, through a judicious choice of the background metric $g$, to control the sign of $h$.  

\medskip

We prove two type of results in this paper. 
The first one, in Section \ref{NonExi}, establishes a set of sufficient conditions to guarantee the
nonexistence of positive solutions of \eqref{EinLicEqt}. The second one, 
in Section \ref{Exist}, is concerned with the existence of positive solutions of \eqref{EinLicEqt}. 
Our existence result corresponds to (but generalizes) the case of initial data with a positive 
Yamabe--scalar field conformal invariant considered in \cite{CIP}.  More specifically the results
presented here should be contrasted with the partial results indicated in the third row of Table 2
of \cite{CIP}, and specifically with Theorems 4 and 5 in \S 5.4--5.5 of \cite{CIP}.  The results presented 
here apply, for example, when considering initial data for the Einstein--massive--Klein--Gordon system with 
small (relative to the mass), or zero, values of the mean curvature.  The basic variational method  employed here 
is to use the mountain pass lemma \cite{AR, Rab} to solve a family of
$\varepsilon$-approximated equations, and let then $\varepsilon \to 0$ to obtain a solution of \eqref{EinLicEqt}. 
Finally, Section \ref{EinMaxTh} contains a brief discussion of a class of slightly more general equations  which arise 
when considering the Einstein--Maxwell--scalar field theory.

\section{Nonnexistence of smooth positive solutions}\label{NonExi}

Examples of nonexistence results involving pointwise 
conditions on $h$, $A$, and $B$ are easy to get. Let $u$ be a smooth positive solution of 
\eqref{EinLicEqt}, and $x_0$ be a point where $u$ is minimum.
Then $\Delta_gu(x_0) \le 0$ and we get that $h(x_0)u(x_0) \ge B(x_0)u(x_0)^{2^\star-1} + A(x_0)u(x_0)^{-2^\star -1}$. 
Let us assume that both $A$ and $B$ are positive functions. We have
\begin{equation}\label{eqf:2}
h(x_0)  \ge B(x_0) \, X + A(x_0)  \, X^{1-n}\hskip.1cm ,
\end{equation}
where we have set $X = u(x_0)^{\frac{4}{n-2}} $. Studying the least value of the right hand side of (\ref{eqf:2}) (considered as a function of $X$), we get that \eqref{EinLicEqt} does not possess 
a smooth positive solution if
\begin{equation}\label{NecPtwCdt}
\frac{n^n}{(n-1)^{n-1}}  >  \max_M  \left( \frac{(h^+)^n}{A \, B^{n-1}} \right)  \hskip.1cm .
\end{equation}
It also follows from \eqref{eqf:2} that 
$$u(x) \ge u (x_0) \ge \left(\min_M\frac{A}{ h^+ }\right)^{\frac{n-2}{4(n-1)}}$$
for all $x \in M$.  The idea of getting such a bound  will be used again in Section \ref{Exist} when proving Theorem \ref{ExiThm}. 
We now obtain a nonexistence result involving the Lebesgue norm of the functions $A,B$ and $h$. 

\begin{thm}\label{NonExiThm0} Let $(M,g)$ be a smooth compact Riemannian manifold 
of dimension $n \ge 3$. Let also $h$, $A$, and $B$ be smooth functions on $M$ with $A \ge 0$ in $M$. If $B > 0$ in $M$, and 
\begin{equation}\label{NecCdt1}
\left( \frac{n^n}{(n-1)^{n-1}} \right)^{\frac{n+2}{4n}} \,   \int_M A^{\frac{n+2}{4n}} \, B^{\frac{3n-2}{4n}} \, dv_g  
>   \int_M  (h^+)^{\frac{n+2}{4}}  \, B^{\frac{2-n}{4}} \,  dv_g 
\hskip.1cm ,
\end{equation}
where $h ^+ = \max(0,h)$, 
then the Einstein--scalar field Lichnerowicz equation 
\eqref{EinLicEqt} does not possess any smooth positive solution.
\end{thm}

\begin{proof} We assume first that $B > 0$. Let $u$ be a smooth positive solution of \eqref{EinLicEqt}. Integrating 
\eqref{EinLicEqt} over $M$ we get that
\begin{equation}\label{PrThm1Eqt1}
\int_MBu^{2^\star-1}dv_g + \int_M\frac{Adv_g}{u^{2^\star +1}} = \int_Mhudv_g\hskip.1cm .
\end{equation}
By H\"older's inequality,
$$\int_Mhudv_g \le \left(\int_M  (h^+)^{\frac{n+2}{4}}  \, B^{\frac{2-n}{4}}  \, dv_g\right)^{\frac{4}{n+2}} 
\left(\int_M B \, u^{2^\star-1}dv_g\right)^{\frac{n-2}{n+2}}\hskip.1cm .$$
Again by using H\"older's inequality, 
$$ \int_M A^{\frac{n+2}{4n}} \, B^{\frac{3n-2}{4n}} \, dv_g  \le  \left(   \int_M B \, u^{2^\star-1}dv_g \right)^{\frac{3n-2}{4n}} \, \left( \int_M\frac{Adv_g}{u^{2^\star +1}}\right)^{\frac{n+2}{4n}}
\hskip.1cm .$$
Collecting these inequalities and using (\ref{PrThm1Eqt1}), we get 
\begin{equation}
X + \left( \int_M A^{\frac{n+2}{4n}} \, B^{\frac{3n-2}{4n}} \, dv_g\right)^{\frac{4n}{n+2}} \, X^{1-n}Ê\leq  \left(\int_M (h^+)^{\frac{n+2}{4}}  \, B^{\frac{2-n}{4}} \, dv_g\right)^{\frac{4}{n+2}}
\label{eqf:1}
\end{equation}
where we have set 
$$X = \left( \int_M B \, u^{2^\star-1}dv_g \right)^{\frac{4}{n+2}}$$
The study of the minimal value of the function of $X$ which appears on the left hand side of (\ref{eqf:1}) implies that 
$$\frac{n^n}{(n-1)^{n-1}}  \,\left( \int_M A^{\frac{n+2}{4n}} \, B^{\frac{3n-2}{4n}} \, dv_g \right)^{\frac{4n}{n+2}} 
\leq  \left( \int_M (h^+)^{\frac{n+2}{4}}  \, B^{\frac{2-n}{4}} \,  \, dv_g \right)^{\frac{4n}{n+2}}$$
This completes the proof of the theorem. 
\end{proof}

\medskip
Many more restrictive nonexistence conditions can be obtained easily from (\ref{NecCdt1}). For example, replacing $B$ by 
$\min_MB$ in the two integrals in \eqref{NecCdt1}, we get that if
$$\left(\frac{n^n}{(n-1)^{n-1}}  \right)^{\frac{n+2}{4n}} \int_M A^{\frac{n+2}{4n}} dv_g  >  
\frac{ \int_M (h^+)^{\frac{n+2}{4}} dv_g}{\left(\min_MB\right)^{\frac{(n-1)(n+2)}{4n}}}$$
is fulfilled, then (\ref{NecCdt1}) holds true and the Einstein--scalar field Lichnerowicz equation 
\eqref{EinLicEqt} does not possess any smooth positive solution. 
In the same spirit, note that condition (\ref{NecPtwCdt}) is more restricitive than (\ref{NecCdt1}) since, for any triple of functions satisfying (\ref{NecPtwCdt}) we have 
$$\frac{n^n}{(n-1)^{n-1}} \,  A \, B^{\frac{3n-2}{n+2}} >  (h^+)^n \,  B^{\frac{n(2-n)}{n+2}}$$
raising this to the power $\frac{n+2}{4n}$ and integrating the result over $M$ yields (\ref{NecCdt1}).

\medskip
In what follows we let $S = S(M,g)$, $S > 0$, be the Sobolev constant of $(M, g)$ defined as the smallest $S > 0$ such that
\begin{equation}\label{SobIne}
\int_M\vert u\vert^{2^\star}dv_g  \le S \, \left( \int_M\left(\vert\nabla u\vert^2 + u^2\right)dv_g \right)^{\frac{2^\star}{2}}
\end{equation}
for all $u \in H^1(M)$. Explicit upper bounds for $S$ can be given 
in special geometries, like, see Ilias \cite{Ili}, when the Ricci curvature of the manifold is positive. Concerning lower bounds, 
it is well-known that $S \ge K_n^{2^\star}$, where $K_n$ is the sharp Sobolev constant in the 
$n$-dimensional Euclidean space for the Sobolev inequality $\Vert u\Vert_{L^{2^\star}} \le K_n\Vert\nabla u\Vert_{L^2}$. By letting 
$u = 1$ in \eqref{SobIne} we also get that $S \ge V_g^{-2^\star/n}$, where $V_g$ is the volume of $M$ with respect to $g$. 
Using this, we prove some nonexistence result for solutions with bound an a priori bound on their $H^1$ energy. 

\begin{thm}\label{NonExiThm} Let $(M,g)$ be a smooth compact Riemannian manifold 
of dimension $n \ge 3$. Let also $h$, $A$, and $B$ be smooth functions on $M$ with $A \ge 0$ in $M$.  If $B$ is arbitrary, not 
necessarily positive, and
\begin{equation}\label{NecCdt2}
\int_MA^{\frac{1}{2}}dv_g > S\Lambda^{2^\star}
\left(\max_MB^- + \frac{\max\left(1,\max_Mh^+\right)}{S\Lambda^{\frac{4}{n-2}}}\right)^{\frac{1}{2}}
\end{equation}
for some $\Lambda > 0$, where $B^- = \max(0,-B)$ and $S$ is as in \eqref{SobIne}, 
then the Einstein--scalar field Lichnerowicz equation 
\eqref{EinLicEqt} does not possess smooth positive solutions of energy $\Vert u\Vert_{H^1} \le \Lambda$. Moreover, 
\eqref{NecCdt2} is sharp in the sense that the power $p = \frac{1}{2}$ in the left hand side of 
\eqref{NecCdt2} cannot be improved, and that the bound on the energy cannot be removed.
\end{thm}

\begin{proof} We prove here that  \eqref{NecCdt2} prohibits the existence of positive solutions of \eqref{EinLicEqt}. The discussion on the sharpness of this condition  is postponed after the proof.  

\medskip
Let $u$ be a smooth positive 
solution of \eqref{EinLicEqt} such that $\Vert u\Vert_{H^1} \le \Lambda$, $\Lambda > 0$. Let $C_h = \max\left(1,\max_Mh^+\right)$, 
where $h^+ = \max(0,h)$. Then,
\begin{equation}\label{PrThm1Eqt6}
\int_M\left(\vert\nabla u\vert^2 + hu^2\right)dv_g \le C_h\int_M\left(\vert\nabla u\vert^2+u^2\right)dv_g.
\end{equation}
Multiplying \eqref{EinLicEqt} by $u$, and integrating over $M$, we get by \eqref{PrThm1Eqt6} that
\begin{equation}\label{PrThm1Eqt7}
\int_MBu^{2^\star}dv_g + \int_M\frac{Adv_g}{u^{2^\star}} \le C_h\Lambda^2\hskip.1cm .
\end{equation}
By the Sobolev inequality \eqref{SobIne} we can write that
\begin{equation}\label{PrThm1Eqt8}
\int_MBu^{2^\star}dv_g \ge  -\left(\max_MB^-\right)S\Lambda^{2^\star}\hskip.1cm ,
\end{equation}
where $B^- = \max\left(0,-B\right)$. Then, by combining \eqref{PrThm1Eqt7}--\eqref{PrThm1Eqt8} we get that
\begin{equation}\label{PrThm1Eqt9}
\int_M\frac{Adv_g}{u^{2^\star}} \le C_h\Lambda^2 + \left(\max_MB^-\right)S\Lambda^{2^\star}\hskip.1cm .
\end{equation}
Now, H\"older inequality yields
\begin{equation}\label{PrThm1Eqt4}
\int_MA^{\frac{1}{2}} dv_g \le \left(\int_M\frac{A dv_g}{u^{2^\star}}\right)^{\frac{1}{2}} \left(\int_Mu^{2^\star}dv_g\right)^{\frac{1}{2}} .
\end{equation}
By combining this inequality with \eqref{PrThm1Eqt9}, and 
by the Sobolev inequality \eqref{SobIne}, we get that
$$\int_MA^{\frac{1}{2}}dv_g \le S \Lambda^{2^\star}
\left(\max_MB^- + \frac{C_h}{S \Lambda^{\frac{4}{n-2}}}\right)^{\frac{1}{2}}
\hskip.1cm .$$
This proves the theorem. 
\end{proof}

We now discuss the sharpness of \eqref{NecCdt2} in Theorem \ref{NonExiThm}. 
The Yamabe equation on a Riemannian manifold $(M,g)$ may be written as
\begin{equation}\label{YamEqt}
\Delta_gu + \frac{n-2}{4(n-1)}R(g) u = u^{2^\star-1}\hskip.1cm ,
\end{equation}
where $R(g)$ is the scalar curvature of $g$. A positive solution $u>0$ of \eqref{YamEqt} corresponds to 
a conformally related metric $\tilde g = u^{2^\star-2}g$ with constant positive scalar curvature
$R(\tilde g) = \frac{4(n-1)}{n-2}$.
Now, any solution of \eqref{YamEqt} is 
a solution of \eqref{EinLicEqt} when we let $h = \frac{n-2}{4(n-1)} \, R(g)$, $B = \alpha$, 
and $A = (1-\alpha)u^{22^\star}$ for some $\alpha \in \mathbb{R}$. This provides a transformation rule 
for rewriting equations like 
\eqref{YamEqt} into equations like \eqref{EinLicEqt}. On the unit sphere $(S^n,g)$, for which $R(g) = n(n-1)$, 
we know (see, for instance, Aubin \cite{Aub82}) that 
there exist families $(u_\varepsilon)_\varepsilon$ of solutions of \eqref{YamEqt}, $\varepsilon > 0$, such that 
$\Vert u_\varepsilon\Vert_{H^1} = K_n^{-n}+o(1)$ for all $\varepsilon > 0$, and $\Vert u_\varepsilon\Vert_{L^p} \to +\infty$ 
as $\varepsilon \to 0$ for all $p > 2^\star$, where $K_n$ is the sharp Sobolev constant in the 
$n$-dimensional Euclidean space for the Sobolev inequality 
$\Vert u\Vert_{L^{2^\star}} \le K_n\Vert\nabla u\Vert_{L^2}$.
Letting $\alpha = \frac{1}{2}$, the above transformation rule \eqref{YamEqt}$\rightarrow$\eqref{EinLicEqt} provides 
a family of Einstein--scalar field Lichnerowicz type equations indexed by $\varepsilon > 0$, 
with $h$ and $B$ independent of $\varepsilon$, such that any equation in the family possesses a solution of energy less than 
or equal to $2K_n^{-n}$, and for which $\int_MA_\varepsilon^pdv_g \to +\infty$ as $\varepsilon \to 0$ for all $p > \frac{1}{2}$. This proves that 
the power $p = \frac{1}{2}$ in the left hand side of \eqref{NecCdt2} cannot be improved. This example can be 
modified in different ways with the constructions given in Brendle \cite{Bre} and in Druet and Hebey \cite{DruHeb}.

\medskip
We prove next that the bound on the energy in Theorem \ref{NonExiThm} cannot be removed.
By Druet and Hebey \cite{DruHeb} we know that on the unit sphere in dimension $n \ge 6$, or on any quotient $(M,g)$ of the unit sphere 
in dimension $n \ge 6$, there exist families 
$(h_\varepsilon)_\varepsilon$ of smooth functions, such that $h_\varepsilon \to \frac{n(n-2)}{4}$ in $C^1(M)$, and 
families $(u_\varepsilon)_\varepsilon$ of smooth positive functions such that, for any $\varepsilon > 0$, 
$u_\varepsilon$ solves the Yamabe type equation
\begin{equation}\label{YamEqType}
\Delta_{g_0}u_\varepsilon +h_\varepsilon u_\varepsilon = u_\varepsilon^{2^\star-1}\hskip.1cm ,
\end{equation}
and such that $\Vert u_\varepsilon\Vert_{H^1} \to +\infty$ as 
$\varepsilon \to 0$. Rewriting \eqref{YamEqType} with the transformation rule 
\eqref{YamEqt}$\rightarrow$\eqref{EinLicEqt}, we see that the $u_\varepsilon$'s solve \eqref{EinLicEqt} 
with $h = h_\varepsilon$, $B = \alpha$, 
and $A = (1-\alpha)u_\varepsilon^{22^\star}$ for some $\alpha \in \mathbb{R}$. Letting $\alpha = \frac{1}{2}$, 
we get families of Einstein--scalar field Lichnerowicz type equations indexed by $\varepsilon > 0$ such that any equation 
in the family possesses a solution, $B$ is independent of $\varepsilon$, 
the $h_\varepsilon$'s converge in the $C^1$ -topology to a positive constant function,
and $\int_MA_\varepsilon^{1/2}dv_g \to +\infty$ as $\varepsilon \to 0$. In particular, we cannot hope to get 
that there exists $C = C(n,h,B)$, depending on the manifold and 
continuously on $h$ and $B$ in the $C^0$-topology, like this is the case for the constant 
in \eqref{NecCdt2} when $\Lambda$ is fixed, 
such that if $\int_MA^{1/2}dv_g \ge C$, then the Einstein--scalar field Lichnerowicz type equation \eqref{EinLicEqt} does 
not possess a smooth positive solution. This proves that the bound on the energy in Theorem \ref{NonExiThm} cannot be removed.

\medskip

In the same circle of ideas, we mention that if $B > 0$ in $M$, then we can give another form to \eqref{NecCdt2} where 
the constant appears as $C\Lambda^2$. In order to get this dependancy in $\Lambda^2$ we may proceed 
as in the proof of Theorem \ref{NonExiThm}, but now getting bounds from the estimate \eqref{PrThm1Eqt7}. 
By \eqref{PrThm1Eqt7}, since we assumed that $B > 0$ in $M$, we can write that 
\begin{equation}\label{AddBdsEn}
\int_Mu^{2^\star}dv_g \le \frac{C_h\Lambda^2}{\min_MB}\hskip.2cm\hbox{and}\hskip.2cm 
\int_M\frac{Adv_g}{u^{2^\star}} \le C_h\Lambda^2\hskip.1cm .
\end{equation}
Then, by \eqref{PrThm1Eqt4} as in the proof of the second part of Theorem \ref{NonExiThm}, we get from \eqref{AddBdsEn} that 
\eqref{EinLicEqt} does not possess a smooth positive solution if
\begin{equation}\label{IntCds}
\int_MA^{\frac{1}{2}}dv_g > \frac{\max\left(1,\max_Mh^+\right)\Lambda^2}{\left(\min_MB\right)^{\frac{1}{2}}}\hskip.1cm .
\end{equation}
Condition \eqref{IntCds} is complementary to the condition in Theorem \ref{NonExiThm}. 
For large $\Lambda$'s, \eqref{IntCds} is better than 
\eqref{NecCdt2} since it involves the energy $\Lambda^2$ and not $\Lambda^{2(n-1)/(n-2)}$.

\section{Existence of a smooth positive solution}\label{Exist}

In this section we use the mountain pass lemma \cite{AR, Rab}, to get existence results that complement the nonexistence results presented in 
Theorem \ref{NonExiThm}. More precisely, we  prove that  if $\int_MAdv_g $ is sufficiently small, and $A > 0$ in $M$, then \eqref{EinLicEqt} possesses a solution. When $A\equiv 0$, \eqref{EinLicEqt} is the prescribed scalar curvature equation and we know from  Kazdan and Warner \cite{KazWar} that there are situations in which the equation does not possess  a solution.

\medskip
In the sequel we assume that the function $h$ is chosen so that $\Delta_g+h$ is coercive. This amounts to say that there exists a constant  $K_h = K(M,g,h) > 0$, such that
$$\int_M \vert u \vert^{2} \, dv_g  \le K_h \, \int_M \left(\vert\nabla u\vert^2+h \, u^2\right)dv_g$$
for all $u \in H^1(M)$. It will be convenient to define 
\begin{equation}\label{eqf:5}
\|Êu\|_{H^1_h} =  \left( \int_M \left(\vert\nabla u\vert^2+h \, u^2\right)dv_g \right)^{\frac{1}{2}}.
\end{equation}
We also denote by $S_h = S(M,g,h) > 0$, the Sobolev constant defined to be the smallest constant $S_h >0$ such that
\begin{equation}\label{eqf:6}
 \int_M \vert u \vert^{2^\star } dv_g  \le S_h \, \left( \int_M \left(\vert\nabla u\vert^2+h \, u^2\right)dv_g \right)^{\frac{2^\star}{2}}
\end{equation}
for all $u \in H^1(M)$.  

\medskip
Observe that, if $h > 0$ in $M$, then $\Delta_g + h$ is coercive and conversely coercivity implies that $\int_M hdv_g >  0$, and thus that $\max_M h > 0$. Also observe that if $A, B \ge 0$, $A+B >0$, and if \eqref{EinLicEqt} possesses a smooth positive solution, then $\Delta_g+h$ is coercive. Indeed,  in that case, there exists a function $u > 0$ such that $\Delta_gu + hu > 0$ everywhere in $M$, and the 
existence of such an $u$ implies the coercivity of $\Delta_g + h$.

\medskip
Finally,  as already mentioned, when $h > 0$ in $M$, then $\Delta_g + h$ is coercive and we have the bound
$$S_h \le  \, \max \left(1,\frac{1}{\min_M h} \right)^{\frac{2^\star}{2}} \, S\hskip.1cm .$$
where $S = S(M,g) >0$ is the Sobolev constant defined in (\ref{SobIne}). 

\medskip
We prove here that the following existence result holds true.

\begin{thm}\label{ExiThm} Let $(M,g)$ be a smooth compact Riemannian manifold 
of dimension $n \ge 3$. Let $h$, $A$, and $B$ be smooth functions on $M$ for which 
$\Delta_g+h$ is coercive, $A > 0$ in $M$, and $\max_M B > 0$. 
There exists a constant $C = C(n)$, $C >0$ depending only on $n$, such that if
\begin{equation}\label{SufExiCdt}
\|Ê\varphi  \|_{H^1_h}^{2^\star} \int_M \frac{A}{\varphi^{2^\star}} \, dv_g \le  \frac{C}{Ê( S_h \, \max_M |B|Ê)^{n-1}}
\end{equation}
and
$$\int_M B \varphi^{2^\star} dv_g >0$$
for some smooth positive function $\varphi >0$ in $M$, where $\|\cdot\|_{H^1_h}$ is as in \eqref{eqf:5} 
and $S_h$ is as in \eqref{eqf:6}, 
then the Einstein--scalar field Lichnerowicz equation \eqref{EinLicEqt} possesses a smooth positive solution.
\end{thm}

\begin{proof}{\bf Preliminary computations} We define $I^{(1)} : H^1(M) \to \mathbb{R}$ by
\begin{equation}\label{PrThm2Eqt2}
I^{(1)} (u) = \frac{1}{2}\int_M\left(\vert\nabla u\vert^2+hu^2\right)dv_g - \frac{1}{2^\star}\int_MB(u^+)^{2^\star}dv_g \hskip.1cm ,
\end{equation}
and if we fix $\varepsilon > 0$ we define $I^{(2)}_\varepsilon : H^1(M) \to \mathbb{R}$ by
\begin{equation}\label{PrThm2Eqt2-bis}
I^{(2)}_{\varepsilon} (u) = \frac{1}{2^\star}\int_M\frac{Adv_g}{(\varepsilon +(u^+)^2)^{2^\flat}}\hskip.1cm ,
\end{equation}
where 
$$2^\flat = \frac{2^\star}{2}\hskip.1cm .$$
Obviously, for any $u \in H^1(M)$ we can write 
\begin{equation}\label{PrThm2Eqt3}
\Phi (\Vert u\Vert_{H^1_h})  \le I^{(1)} (u)  \le  \Psi (\Vert u\Vert_{H^1_h})
\end{equation}
if the functions  $\Phi, \Psi : [0,+\infty) \to \mathbb{R}$ are defined by
\begin{equation}\label{PrThm2Eqt4}
\Phi(t) = \frac{1}{2}t^2 - \frac{\max_M |B|}{2^\star} S_h \, t^{2^\star}
\end{equation}
and 
\begin{equation}\label{PrThm2Eqt7}
\Psi(t) =\frac{1}{2}t^2 + \frac{\max_M |B|}{2^\star}S_h \, t^{2^\star}
\end{equation}
for $t \in \mathbb{R}$, where $S_h > 0$ and $\Vert  \, \cdot \, \Vert_{H^1_h}$ are as in \eqref{eqf:5} and \eqref{eqf:6}. 

\medskip
Let $t_0 > 0$ be given by
\begin{equation}\label{PrThm2Eqt5}
t_0 = \left(\frac{1}{S_h \, \max_M |B|}\right)^{\frac{n-2}{4}}
\end{equation}
so that $\Phi$ is increasing in $[0,t_0]$, and decreasing in $[t_0,+\infty)$. We define $\theta >0$ such that
$$\theta^2  = \frac{1}{2(n-1)}$$
and $t_1 = \theta \, t_0$ for $t_0$ as in \eqref{PrThm2Eqt5}. It is easy to check that  
\begin{equation}\label{AddEqtFeb1}
\Psi (t_1) \leq \theta^2  \, \frac{2^\star +2}{2^\star -2} \,  \Phi (t_0) \leq  \frac{1}{2}Ê\, \Phi (t_0)
\hskip.1cm ,
\end{equation}
where $\Phi$ and $\Psi$ are as in \eqref{PrThm2Eqt4} and \eqref{PrThm2Eqt7}. 
Finally, we define the functional
\begin{equation}\label{DefFunctFeb}
I_\varepsilon  = I^{(1)} + I^{(2)}_\varepsilon\hskip.1cm ,
\end{equation}
where $I^{(1)}$ and $I^{(2)}_\varepsilon$ are as in \eqref{PrThm2Eqt2} and \eqref{PrThm2Eqt2-bis}. 
Let $\varphi \in C^\infty(M)$, $\varphi > 0$ in $M$, be the function in the statement of the theorem. In particular  
\begin{equation}\label{PrThm2Eqt1}
\int_M B \varphi^{2^\star}dv_g > 0 \hskip.1cm .
\end{equation}
and, without loss of generality, we can assume that 
$$\Vert\varphi\Vert_{H^1_h} = 1\hskip.1cm .$$
Now, provided the constant $C$ in \eqref{SufExiCdt} is chosen to be 
$$C =\theta^{2^\star} \frac{2^\star -2}{4}\hskip.1cm ,$$
we find that (\ref{SufExiCdt}) precisely translates into
\begin{equation}\label{AddEqtFeb2}
\frac{1}{2^\star} \, \int_M \frac{A}{ (t_1 \, \varphi)^{2^\star}} \, dv_g \leq  \frac{1}{2}  \,  \Phi(t_0)
\end{equation}
and by \eqref{PrThm2Eqt3}, \eqref{AddEqtFeb1}, and \eqref{AddEqtFeb2} we get that 
\begin{equation}
I_\varepsilon (t_1 \, \varphi )  \leq  \Phi ( t_0) <  I_\varepsilon (t_0 \, \varphi )
\label{PrThm2Eqt12}
\end{equation}
Finally, \eqref{PrThm2Eqt1} implies that
$$\lim_{+\infty} I_\varepsilon (t \, \varphi ) = - \infty\hskip.1cm .$$
Hence we can choose  $t_2 >t_0$ such that 
\begin{equation}
I_\varepsilon (t_2 \, \varphi )  <0\hskip.1cm ,
\label{PrThm2Eqt11}
\end{equation}
where $I_\varepsilon$ is the functional in \eqref{DefFunctFeb}.

\medskip\noindent
{\bf Application of the Mountain Pass Lemma} By \eqref{PrThm2Eqt12} and 
\eqref{PrThm2Eqt11}, we can apply the mountain pass lemma \cite{AR, Rab} to the functional $ I_\varepsilon$.
Let
\begin{equation}\label{PrThm2Eqt14}
c_\varepsilon  = \inf_{\gamma \in \Gamma}\max_{u \in \gamma}I_\varepsilon (u)\hskip.1cm ,
\end{equation}
where $\Gamma$ stands for the set of continuous paths joining $u_1 = t_1 \varphi $ to $u_2 = t_2 \varphi$.  Observe that 
$c_\varepsilon > \Phi (t_0)$ and, taking the path $\gamma (t) =  t \, \varphi$,  for $t \in [t_1, t_2]$, we see that  $c_\varepsilon$ is bounded uniformly as $\varepsilon$ tends to $0$. We will keep in mind, for further use that 
\begin{equation}
\Phi(t_0) < c_\varepsilon \le c
\label{PrThm2Eqt16}
\end{equation}
for all $\varepsilon$ small enough, where $c > 0$ is independent of $\varepsilon$. 

\medskip
By the mountain pass lemma we get that there exists a sequence $(u_k)_k$ in $H^1(M)$ such that
\begin{equation}\label{PrThm2Eqt15}
I_\varepsilon (u_k) \to c_\varepsilon \hskip.2cm\hbox{and}\hskip.2cm I_\varepsilon^\prime(u_k) \to 0
\end{equation}
as $k \to +\infty$. 

By \eqref{PrThm2Eqt15}, 
\begin{equation}\label{PrThm2Eqt17}
\begin{split}
&\int_M(\nabla u_k\nabla\varphi)dv_g + \int_Mhu_k\varphi dv_g - \int_MB(u_k^+)^{2^\star-1}\varphi dv_g\\
&= \int_M\frac{Au_k^+\varphi dv_g}{(\varepsilon + (u_k^+)^2)^{2^\flat+1}} + o\left(\Vert\varphi\Vert_{H^1_h}\right)
\end{split}
\end{equation}
for all $\varphi \in H^1(M)$, where $(\nabla u_k\nabla\varphi)$ stands for the pointwise scalar product 
of $\nabla u_k$ and $\nabla\varphi$ with respect to $g$, and
\begin{equation}\label{PrThm2Eqt18}
\begin{split}
&\frac{1}{2}\int_M\left(\vert\nabla u_k\vert^2+hu_k^2\right)dv_g - \frac{1}{2^\star}\int_MB(u_k^+)^{2^\star}dv_g\\
&+ \frac{1}{2^\star}\int_M\frac{Adv_g}{(\varepsilon + (u_k^+)^2)^{2^\flat}} = c_\varepsilon + o(1)\hskip.1cm .
\end{split}
\end{equation}
Combining \eqref{PrThm2Eqt17} with $\varphi = u_k$, and \eqref{PrThm2Eqt18}, we get that
\begin{equation}\label{PrThm2Eqt19}
\begin{split}
&\frac{1}{n}\int_MB(u_k^+)^{2^\star}dv_g + \frac{1}{2}\int_M\frac{A(u_k^+)^2dv_g}{(\varepsilon + (u_k^+)^2)^{2^\flat +1}}\\
&+ \frac{1}{2^\star}\int_M\frac{Adv_g}{(\varepsilon + (u_k^+)^2)^{2^\flat}} = 
c_\varepsilon + o\left(\Vert u_k\Vert_{H^1_h}\right) + o(1)\hskip.1cm ,
\end{split}
\end{equation}
and it follows from \eqref{PrThm2Eqt19} that for $k$ sufficiently large,
\begin{equation}\label{PrThm2Eqt20}
\frac{1}{n}\int_MB(u_k^+)^{2^\star}dv_g \le 2c_\varepsilon + o\left(\Vert u_k\Vert_{H^1_h}\right)
\hskip.1cm .
\end{equation}
By \eqref{PrThm2Eqt18} and \eqref{PrThm2Eqt20} we then get that 
for $k$ sufficiently large
\begin{equation}\label{PrThm2Eqt21}
\begin{split}
\int_M\left(\vert\nabla u_k\vert^2+ h u_k^2\right)dv_g
&\le \frac{n-2}{n}\int_MB(u_k^+)^{2^\star}dv_g +  4 \, c_\varepsilon \\
&\le 2n \, c_\varepsilon + o\left(\Vert u_k\Vert_{H^1_h}\right)\hskip.1cm .
\end{split}
\end{equation}
In particular, by \eqref{PrThm2Eqt20} and \eqref{PrThm2Eqt21},
\begin{equation}\label{PrThm2Eqt22}
\begin{split}
&\int_M\left(\vert\nabla u_k\vert^2+ h u_k^2\right)dv_g \le  2n \, c_\varepsilon + 1
\hskip.1cm ,\hskip.1cm\hbox{and}\\
&-\frac{4n}{n-2}c_\varepsilon \le \int_MB(u_k^+)^{2^\star}dv_g \le 3nc_\varepsilon
\end{split}
\end{equation}
for $k$ sufficiently large, where $c_\varepsilon$ is as in \eqref{PrThm2Eqt14}. By \eqref{PrThm2Eqt22}, the 
sequence $(u_k)_k$ is bounded in $H^1(M)$. Up to passing to a subsequence we may then assume that 
there exists $u_\varepsilon \in H^1(M)$ such that $u_k \rightharpoonup u_\varepsilon$ weakly in $H^1(M)$, 
$u_k \rightarrow u_\varepsilon$ strongly in $L^p(M)$ for some 
$p > 2$, and $u_k \rightarrow u_\varepsilon$ almost everywhere 
in $M$ as $k\to +\infty$. As a consequence,
\begin{equation}\label{PrThm2Eqt23}
\begin{split}
&(u_k^+)^{2^\star-1} \rightharpoonup (u_\varepsilon^+)^{2^\star-1}
\hskip.1cm\hbox{weakly in}\hskip.1cm L^{2^\star/(2^\star-1)}(M)\hskip.1cm ,\hskip.1cm\hbox{and}\\
&\frac{u_k^+}{(\varepsilon + (u_k^+)^2)^q}\rightarrow 
\frac{u_\varepsilon^+}{(\varepsilon + (u_\varepsilon^+)^2)^q}
\hskip.1cm\hbox{strongly in}\hskip.1cm L^2(M)
\end{split}
\end{equation}
for all $q > 0$, 
as $k \to +\infty$. Indeed, by \eqref{PrThm2Eqt22}, the $(u_k^+)^{2^\star-1}$'s are bounded in $L^{2^\star/(2^\star-1)}(M)$. 
Since they converge almost everywhere to $(u_\varepsilon^+)^{2^\star-1}$, the first equation in \eqref{PrThm2Eqt23} follows 
from standard integration theory. 
By the Lebesgue's dominated convergence theorem we also have that 
$(\varepsilon + (u_k^+)^2)^{-q} \to (\varepsilon + (u_\varepsilon^+)^2)^{-q}$ strongly in $L^p(M)$ for all 
$p \ge 1$ and all $q > 0$, and since $u_k \to u_\varepsilon$ in $L^p(M)$ for some $p > 2$, we easily get that the second 
equation in \eqref{PrThm2Eqt23} holds true. By \eqref{PrThm2Eqt23}, letting $k \to +\infty$ 
in \eqref{PrThm2Eqt17}, it follows that $u_\varepsilon$ satisfies
\begin{equation}\label{PrThm2Eqt24}
\Delta_gu_\varepsilon + hu_\varepsilon = B(u_\varepsilon^+)^{2^\star-1} + 
\frac{Au_\varepsilon^+}{(\varepsilon + (u_\varepsilon^+)^2)^{2^\flat+1}}
\end{equation}
in the weak sense.
The weak maximum principle and \eqref{PrThm2Eqt24} imply that $u_\varepsilon \ge 0$. As a consequence,
\begin{equation}\label{PrThm2Eqt25}
\Delta_gu_\varepsilon + hu_\varepsilon = Bu_\varepsilon^{2^\star-1} + 
\frac{Au_\varepsilon}{(\varepsilon + u_\varepsilon^2)^{2^\flat+1}}
\end{equation}
in the weak sense. 

\medskip\noindent
{\bf Regularity and positivity of the solution} We may rewrite \eqref{PrThm2Eqt25} as
$$\Delta_gu_\varepsilon + \left(h-\frac{A}{(\varepsilon + u_\varepsilon^2)^{2^\flat+1}}\right)u_\varepsilon 
= Bu_\varepsilon^{2^\star-1}\hskip.1cm ,$$
and since
$$h - \frac{A}{(\varepsilon + u_\varepsilon^2)^{2^\flat+1}} \in L^\infty(M)\hskip.1cm ,$$
the regularity arguments developed in Trudinger \cite{Tru} 
apply to \eqref{PrThm2Eqt25}. It follows that $u_\varepsilon \in L^s(M)$ 
for some $s > 2^\star$. Since we have that $A(\varepsilon + u_\varepsilon^2)^{-{2^\flat+1}}u_\varepsilon \in L^p(M)$ 
if $u_\varepsilon \in L^p(M)$, and $u_\varepsilon \in L^s(M)$ for some $s > 2^\star$, the standard 
bootstrap procedure, together with regularity theory, gives that $u_\varepsilon \in H^{2,p}(M)$ for all $p \ge 1$, where 
$H^{2,p}$ is the Sobolev space of functions in $L^p$ with two derivatives in $L^p$.  By the Sobolev 
embedding theorem we then get that the right hand side in \eqref{PrThm2Eqt25} is in $C^{0,\alpha}(M)$ 
for $\alpha \in (0,1)$, and by regularity theory it follows that $u_\varepsilon \in C^{2,\alpha}(M)$ for 
$\alpha \in (0,1)$. In particular, the strong maximum principle can be applied and we get that either 
$u_\varepsilon \equiv 0$, or $u_\varepsilon > 0$ in $M$. Then we easily get that $u_\varepsilon \in C^\infty(M)$ is 
smooth. By \eqref{PrThm2Eqt22} and \eqref{PrThm2Eqt23}, letting 
$k \to +\infty$ in \eqref{PrThm2Eqt19}, we get that
\begin{equation}\label{PrThm3Eqt26}
\frac{1}{2^\star}\int_M\frac{Adv_g}{(\varepsilon + u_\varepsilon^2)^{{2^\flat}}} 
\le (2^\star -1) c\hskip.1cm .
\end{equation}
where $c$ is the upper bound for $c_\varepsilon$. If,  for a sequence of $\varepsilon_j$ tending to $0$,  $u_{\varepsilon_j}$ where to be equal to $0$, we would conclude that 
\begin{equation}\label{PrThm3Eqt27}
\frac{1}{2^\star(2^\star-1)\varepsilon_j^{{2^\flat}}}\int_MAdv_g \leq c 
\end{equation}
which is clearly impossible since we have assumed that $A >0$.  Therefore, for $\varepsilon$ small enough $u_\varepsilon \not\equiv 0$. 
Then, according to the above  discussion, $u_\varepsilon$ is a smooth positive solution of \eqref{PrThm2Eqt25}. By \eqref{PrThm2Eqt22}, and standard 
properties of the weak limit, we also get that
\begin{equation}\label{PrThm3Eqt28}
\int_M\left(\vert\nabla u_\varepsilon\vert^2+ h u_\varepsilon^2\right)dv_g \le  2n c_\varepsilon + 1
\end{equation}
for all $\varepsilon > 0$ small enough. 

\medskip\noindent
{\bf Passing to the limit as $\varepsilon$ tends to $0$} In what follows we let 
$(\varepsilon_k)_k$ be a sequence of positive real numbers such that $\varepsilon_k \to 0$ as $k \to +\infty$ 
and \eqref{PrThm3Eqt27} holds true with $\varepsilon = \varepsilon_k$ for all $k$, and let $u_k = u_{\varepsilon_k}$. 
Then $u_k$ is a smooth positive function in $M$ such that
\begin{equation}\label{PrThm2Eqt25Bis}
\Delta_gu_k + hu_k = Bu_k^{2^\star-1} + 
\frac{Au_k}{(\varepsilon_k + u_k^2)^{2^\flat+1}}
\end{equation}
in $M$ while, by \eqref{PrThm2Eqt16} and \eqref{PrThm3Eqt28},  the sequence $(u_k)_k$ is bounded in $H^1(M)$. Let $x_k$ 
be a point where $u_k$ is minimum. Then $\Delta_gu_k(x_k) \le 0$ and we get with \eqref{PrThm2Eqt25Bis} that
\begin{equation}\label{PrThm2Eqt26Bis}
h(x_k) + |B| (x_k)u_k(x_k)^{2^\star-2} \ge 
\frac{A(x_k)}{(\varepsilon_k+u_k(x_k)^2)^{2^\flat+1}}\hskip.1cm .
\end{equation}
Let $\delta_0 > 0$ be such that
$$\delta_0^{2({2^\flat+1})}\left(\max_Mh + (\max_M |B| )\delta_0^{2^\star-2}\right) = \frac{\min_MA}{2}\hskip.1cm .$$
By \eqref{PrThm2Eqt26Bis} we obtain that $u_k(x_k) \ge \delta_0$, and thus that
\begin{equation}\label{PrThm2Eqt27Bis}
\min_Mu_k \ge \delta_0
\end{equation}
when $k$ is sufficiently large. Since $(u_k)_k$ is bounded in $H^1(M)$ we may assume that there exists 
$u \in H^1(M)$ such that, up to passing to a subsequence, 
$u_k \rightharpoonup u$ weakly in $H^1(M)$, 
$u_k \rightarrow u$ strongly in $L^p(M)$ for some 
$p > 2$, and $u_k \rightarrow u$ almost everywhere 
in $M$ as $k\to +\infty$. By \eqref{PrThm2Eqt27Bis}, $u \ge \delta_0$ almost everywhere in $M$. Still by 
\eqref{PrThm2Eqt27Bis}, we get with 
similar arguments to those used to prove \eqref{PrThm2Eqt23} that
\begin{equation}\label{PrThm2Eqt28Bis}
\begin{split}
&u_k^{2^\star-1} \rightharpoonup u^{2^\star-1}
\hskip.1cm\hbox{weakly in}\hskip.1cm L^{2^\star/(2^\star-1)}(M)\hskip.1cm ,\hskip.1cm\hbox{and}\\
&\frac{u_k}{(\varepsilon_k + u_k^2)^{2^\flat+1}}\rightarrow 
\frac{1}{u^{2^\star +1}}
\hskip.1cm\hbox{strongly in}\hskip.1cm L^2(M)
\end{split}
\end{equation}
as $k\to +\infty$. By \eqref{PrThm2Eqt25Bis} and \eqref{PrThm2Eqt28Bis}, letting $k \to +\infty$ 
in \eqref{PrThm2Eqt25Bis}, we get that $u$ is a 
weak solution of the Einstein--scalar field Lichnerowicz equation \eqref{EinLicEqt}. Rewriting 
\eqref{EinLicEqt} as
$$\Delta_gu + \left(h-\frac{A}{u^{2^\star+2}}\right)u 
= Bu^{2^\star-1}\hskip.1cm ,$$
and since $h - Au^{-2^\star-2} \in L^\infty(M)$, 
the regularity arguments developed in Trudinger \cite{Tru} 
apply to \eqref{EinLicEqt}. It follows that $u \in L^s(M)$ 
for some $s > 2^\star$. Since $u \ge \delta_0$ almost everywhere, and $\delta_0 > 0$, the 
standard bootstrap procedure, together with regularity theory, gives that $u$ is a smooth positive solution of \eqref{EinLicEqt}. 
This ends the proof of the theorem.
\end{proof}

As a remark, the above proof provides an explicit expression for the dimensional 
constant $C$ in \eqref{SufExiCdt}. As another remark, it can be noted that when 
$\int_M B dv_g > 0$, then we can take $\varphi$ to be constant in \eqref{PrThm2Eqt1}. 
In particular, our existence result has the following Corollary.

\begin{cor} Let $(M,g)$ be a smooth compact Riemannian manifold  of dimension $n \ge 3$ and $h$ a smooth functions on $M$ for which 
$\Delta_g+h$ is coercive. There exists a constant $C =C(n,h)$, $C >0$, such that if $A$ and $B$ are smooth functions on $M$, with $A > 0$ in $M$, $\max_M B > 0$, and $\int_M B dv_g > 0$, and  if we further assume that 
\begin{equation}\label{SufExiCdt2}
(\max_M |B|Ê)^{n-1} \,  \int_M A \, dv_g  \le  C(n,h)  \hskip.1cm ,
\end{equation}
then the Einstein--scalar field Lichnerowicz equation \eqref{EinLicEqt} possesses a smooth positive solution.
\end{cor}

When $A>0$ and $B > 0$, we can also take $\varphi = A^{\frac{n-2}{4n}}$ in \eqref{PrThm2Eqt1}, and our existence result has the following Corollary.

\begin{cor} Let $(M,g)$ be a smooth compact Riemannian manifold  of dimension $n \ge 3$ and $h$ a smooth functions on $M$ for which 
$\Delta_g+h$ is coercive. There exists a constant $C =C(n,h)$, $C >0$, such that if $A$ and $B$ are smooth functions on $M$, with $A > 0$ and $B >0$  in $M$ and  if we further assume that 
\begin{equation}\label{SufExiCdt3}
(\max_M |B|Ê)^{n-1} \,  \| A^{\frac{n-2}{4n}} \|_{H^1}^{2^\star} \,  \int_M A^{\frac{1}{2}} \, dv_g  \le  C(n,h)  \hskip.1cm ,
\end{equation}
then the Einstein--scalar field Lichnerowicz equation \eqref{EinLicEqt} possesses a smooth positive solution.
\end{cor}
Interestingly,  Sobolev embedding implies that 
$$\int_M A^{\frac{1}{2}}Ê\, dv_g \leq S \| A^{\frac{n-2}{4n}} \|_{H^1}^{2^\star}$$
and so, if $A$ and $B$ satisfy (\ref{SufExiCdt3}), then 
$$(\max_M |B|Ê)^{n-1} \,  \left( \int_M A^{\frac{1}{2}} \, dv_g  \right)^2  \le  \frac{C(n,h)}{S}  \hskip.1cm ,$$
which is reminiscent of the condition (with the opposite inequality) that ensured the non existence of a solution, which was obtained in Theorem~\ref{NonExiThm}. 

\section{Einstein-Maxwell-scalar field theory}\label{EinMaxTh}

The methods employed in sections \ref{NonExi} and \ref{Exist} are strong enough to deal with additional 
nonlinear negative power terms 
in the equation of the form $Cu^{-p}$ for $C \ge 0$ and $p > 1$. 
Such terms arise, for example, in 
the Einstein-Maxwell-scalar field theory.
Given $(M,g)$ compact of dimension $n \ge 3$, we let $h$, $A$, $B$, and $C$ be smooth functions 
in $M$, and we briefly discuss in this section equations of the form
\begin{equation}\label{EinMaxEqt}
\Delta_gu+hu = Bu^{2^\star-1} + \frac{A}{u^{2^\star +1}} + \frac{C}{u^p}\hskip.1cm ,
\end{equation}
where $A, C \ge 0$ and $p > 1$. In the case of the Einstein-Maxwell-scalar field theory in (spatial) dimension 
$n=3$ we have $p = 3$ and $C \ge 0$ represents the sum of the squares of the norms of the electric and magnetic fields 
on $M$.  The approach we used to prove Theorem \ref{NonExiThm} deals with inequalities  resulting from the signs of the coefficients and the powers  of the unknown function $u$
 and thus applies 
to \eqref{EinMaxEqt}. Let $\hat p = \frac{2^\star+p-1}{2^\star-1}$. Then, if we concentrate on getting nonexistence results of 
smooth positive solutions with no a priori bound 
on the energy, the approach we used to prove Theorem \ref{NonExiThm} gives 
in particular that \eqref{EinMaxEqt} does not possess 
a smooth positive solution if $B > 0$ in $M$, $A, C \ge 0$ in $M$, and either \eqref{NecCdt1} holds true, or
\begin{equation}\label{NonExiCdtEinMax}
\left( \frac{(\alpha+1)^{\alpha+1}}{\alpha^\alpha}\right)^{\frac{1}{\hat p}} 
\int_M C^{\frac{1}{\hat p}} \, B^{\frac{\hat p - 1}{\hat p}} \, dv_g  
>   \int_M  (h^+)^{\frac{n+2}{4}}  \, B^{\frac{2-n}{4}} \,  dv_g 
\hskip.1cm ,
\end{equation}
where $\alpha = (n-2)(p+1)/4$. 
We also do get similar conditions to \eqref{NonExiCdtEinMax} 
for the nonexistence of solutions of \eqref{EinMaxEqt} of energy bounded by $\Lambda$. 
The method we used to prove Theorem \ref{ExiThm} applies to \eqref{EinMaxEqt} as well. Assume $\Delta_g+h$ is coercive, 
$A, C \ge 0$ in $M$, $A + C > 0$ in $M$, and $\max_MB > 0$. 
Following the proof of Theorem \ref{ExiThm} we get that
that there exists $\Lambda = \Lambda(n,p)$, 
$\Lambda > 0$ depending only on $n$ and $p$, such that if
\begin{equation}\label{SufExiCdtTer}
\int_M \frac{A}{\varphi^{2^\star}} \, dv_g 
\le  \frac{\Lambda}{Ê( S_h \, \max_M |B|Ê)^{n-1}}\hskip.2cm ,\hskip.2cm 
\int_M \frac{C}{\varphi^{p-1}} \, dv_g 
\le  \frac{\Lambda}{Ê( S_h \, \max_M |B|Ê)^\alpha}
\end{equation}
and
$$\int_M B \varphi^{2^\star} dv_g >0$$
for some smooth positive function $\varphi >0$ in $M$ such that $\|Ê\varphi  \|_{H^1_h} = 1$, 
where $\|\cdot\|_{H^1_h}$ is as in \eqref{eqf:5}, $S_h$ is as in \eqref{eqf:6}, and $\alpha$ is 
as in \eqref{NonExiCdtEinMax}, then 
\eqref{EinMaxEqt} possesses a smooth positive solution. As for \eqref{SufExiCdt}, 
the constant $\Lambda$ in \eqref{SufExiCdtTer} can be made explicit.

\end{document}